\documentstyle[preprint,aps]{revtex}
\tighten
\begin{document}
\preprint{UAB--FT--354, gr-qc/9412001}
\draft

\title{Perturbative Metric of Charged Black Holes in Quadratic Gravity}

\author{M.Campanelli,}
\address{Fakult\"at f\"ur Physik der Universit\"at Konstanz,
Postfach 5560 M 674, D - 78434 Konstanz, Germany}
\author{
C.O.Lousto\thanks{Permanent Address: Instituto de Astronom\'\i a y
F\'\i sica del Espacio, Casilla de Correo 67 - Sucursal 28, 1428
Buenos Aires,
Argentina. E-mail: lousto@iafe.uba.ar},}
\address{IFAE - Grupo de F\'\i sica Te\'orica, Universidad Aut\'onoma de
Barcelona, E-08193 Bellaterra (Barcelona), Spain. E-mail:
lousto@ifae.es}
\author{and J.Audretsch}
\address{Fakult\"at f\"ur Physik der Universit\"at Konstanz,
Postfach 5560 M 673, D - 78434 Konstanz, Germany}
\date{\today}
\maketitle

\begin{abstract}
We consider perturbative solutions to the classical field equations
coming from a quadratic gravitational lagrangian in four dimensions.
We study the charged, spherically symmetric black hole and explicitly
give corrections up to third order (in the coupling constant $\beta$
multiplying the $R_{\mu\nu}R^{\mu\nu}$ term)
to the Reissner--Nordstr\"om hole metric.
We consider the thermodynamics
of such black holes, in particular, we compute explicitly
its temperature and entropy--area
relation which deviates from the usual $S=A/4$ expression.
\end{abstract}


\section{Introduction}

Quadratic or fourth-order gravity appears to be a good candidate to
represent the low energy limit of the, yet unknown,
quantum theory of gravity.
In fact, it arises as the necessary counterterms one encounters when
trying to renormalize semiclassical theory at one-loop level\cite{BD82},
and as the low-energy limit of string theory (ignoring the dilaton and
antisymmetric fields)\cite{GSW87}.
Consequently, it is interesting to study the classical
solutions of such effective theory. In Ref.\cite{CLA94}
we have developed a method to solve the field equations of the quadratic
gravitational theory in four dimensions
coupled to matter (see also Refs.\cite{PS93} where
a closely related approach was independently derived from a formal
perturbation theory).
The quadratic terms are written
as a function of the matter stress tensor and its derivatives in such
a way to have, order by order, Einsteinian field equations with
an effective $T_{\mu\nu}$ as source.
By successive perturbations around a solution to Einstein Gravity,
which for us represent the zeroth order, one can build up approximate
solutions. In Ref.\cite{CLA94},
we applied this perturbative procedure to find first order solutions
in the coupling constant $\beta$ for the charged black hole
(where in this particular case $\alpha$ contributions vanish).

We consider the following Lagrangian
formulation of quadratic theories
\begin{eqnarray}
I=I_{G}+I_{m}=&&{1\over 16\pi}\int d^4x\sqrt{-g}\bigg\{-2\Lambda + R \cr
&&+ \alpha R^2 + \beta R_{\mu\nu}R^{\mu\nu} + 16\pi{\cal L}_m\bigg\}~,
\label{1.1}
\end{eqnarray}
where we have set $G=1$ and $c=1$ for simplicity.

For the perturbative approach to properly work we consider only small
curvatures, such that
\begin{equation}
\alpha |R|\ll 1 ~~~~,~~~~|\beta R_{\mu\nu}|\ll 1~. \label{2.2}
\end{equation}
{}~~The coupling constants $\alpha$ and $\beta$ are expected to be of
the order of the Planck length squared. They already have upper bounds
of the atomic size order from observations
in the solar system, binary pulsars and cosmology.

It is worth to stress that the coupling constants $\alpha$ and $\beta$
must fulfill the, so called, no-tachyon constraints
\begin{equation}
3\alpha + \beta\geq 0~~~~,~~~~  \beta\leq 0~, \label{2.1}
\end{equation}
that can be deduced upon linearization and asking for a real
mass for both the scalar field $\phi$ related to $R$
and the spin-two field $\psi_{\mu\nu}$ related to $R_{\mu\nu}$
(see Ref. \cite{AEL93} for further details).

The field equations derived by extremizing the action $I$ can be
rewritten to the {\it n-th order} approximation as\cite{CLA94}
\begin{eqnarray}
R_{\mu\nu}^{(n)}&&-{1\over 2}R^{(n)}g^{(n)}_{\mu\nu}+\Lambda
g^{(n)}_{\mu\nu}=8\pi T_{\mu\nu}^{{\rm eff}(n)} =
8\pi T_{\mu\nu}(g_{\mu\nu}^{(n)})\cr\cr
&&-\alpha H_{\mu\nu}(g_{\mu\nu}^{(n-1)}, T_{\mu\nu}^{(n-1)})
-\beta I_{\mu\nu}(g_{\mu\nu}^{(n-1)}, T_{\mu\nu}^{(n-1)})~, \label{2.8}
\end{eqnarray}
where the {\it zeroth-order} corresponds to the ordinary Einstein
equations and \begin{equation}
H_{\mu\nu}=-2R_{;\mu\nu}+2g_{\mu\nu}\Box R-{1\over 2}g_{\mu\nu}R^2
+2R R_{\mu\nu}~, \label{1.2'}
\end{equation}
and
\begin{eqnarray}
I_{\mu\nu}=&&-2R_{\mu~;\nu\alpha}^{\alpha} +\Box R_{\mu\nu}+
{1\over 2}g_{\mu\nu}\Box R\cr
&&+2R_{\mu}^{~\alpha}R_{\alpha\nu}
-{1\over 2}g_{\mu\nu}R_{\alpha\beta}R^{\alpha\beta}~. \label{1.2''}
\end{eqnarray}

In Section II we explicitly compute the effective stress tensor, metric
coefficients and horizon radius to third order in the coupling constant
$\beta$. In Section III we use this results to analyze the thermodynamic
properties of charged black holes in the quadratic theory of gravity.
We then briefly discuss the validity of the laws of thermodynamics for
black holes in
theories of gravitation different from Einstein's general relativity.
Finally, we discuss the relevance of the higher order $\beta$--corrections
in the computations of the extreme black hole case, and in the
``stabilization" of the quadratic solution around the general relativistic
metric.

\section{Charged Black Hole Metric}

We shall now study spherically symmetric solutions to the quadratic
field equations representing charged black holes.
Its metric can be written
(in the Schwarzschild gauge) as
\begin{equation}
ds^2=g_{tt}(r)dt^2+g_{rr}(r)dr^2+r^2d\Omega^2~,    \label{3.19}
\end{equation}
where $d\Omega^2=d\vartheta^2+\sin^2\vartheta d\varphi^2$.
The starting point is the General Relativistic
solution, i.e. the Reissner--Nordstr\"om metric,
$-g_{tt}(r)=g_{rr}(r)^{-1} =(1-2M/r+Q^2/r^2)$.

The non-vanishing components of the energy-momentum tensor are\cite{MTW}
\begin{equation}
T_t^t=T_r^r=-T_{\vartheta}^{\vartheta}=-T_{\varphi}^{\varphi}=
-{Q^2\over 8\pi r^4}~.      \label{3.20}
\end{equation}

The exact metric, solution of the Einstein equations with the effective
source, Eq.\ (\ref{2.8}), can be
written as \cite{LS88} [in the Schwarzschild gauge, Eq.\ (\ref{3.19})]
\begin{equation}
g^{-1}_{rr}(r)=1-{2M\over r}+{8\pi\over r}\int_{\infty}^r
{\tilde r^2 T^{{\rm eff}~t}_t d\tilde r}~,   \label{3.21}
\end{equation}
\begin{equation}
g_{tt}(r)=-g_{rr}(r)^{-1} \exp{\left\{8\pi\int_{\infty}^r{(T_r^r-T_t^t)
^{\rm eff}\tilde r g_{rr}(\tilde r) d\tilde r}\right\}}~.   \label{gtt}
\end{equation}

We use Eq.\ (\ref{3.20}) as the zeroth order effective
$T_{\mu\nu}$
to compute the higher order $T_{\mu\nu}^{\rm eff}$ (Eq.\
(\ref{2.8})) we use a program of analytic manipulation within
``Mathematica"\cite{W91}
to find up to third order corrections to the energy-momentum tensor
\begin{eqnarray}
T_r^{{\rm eff}~r}&=&-{Q^2\over 8\pi r^4}+{4\beta Q^2\over 8\pi r^6}
\bigg(1-{2M\over r}+{Q^2\over r^2}\bigg)\nonumber\\
&&-{\beta^2 Q^2\over 5\pi r^{8}}\bigg(45-220{M\over r}+
280{M^2\over r^2}+135{Q^2\over r^2}-345{MQ^2\over r^3}+
106{Q^4\over r^4}\bigg)
\nonumber\\
&&-{\beta^3 Q^2\over 210\pi r^{10}}\bigg(75600-635040{M\over r}+
1774080{M^2\over r^2}+430080{Q^2\over r^2}-2432640{MQ^2\over r^3}
\nonumber\\
&&-1646400{M^3\over r^3}+847908{Q^4\over r^4}+3385200{M^2Q^2\over r^4}-
2323965{MQ^4\over r^5}+525476{Q^6\over r^6}\bigg)\nonumber\\
&&+{\cal O}(\beta^4)~,\label{trr}
\end{eqnarray}
\begin{eqnarray}
T_t^{{\rm eff}~t}&=&-{Q^2\over 8\pi r^4}+{3\beta Q^2\over 2\pi r^6}
\bigg(1-{2M\over r}+{Q^2\over r^2}\bigg)\nonumber\\
&&-{\beta^2 Q^2\over 5\pi r^{8}}\bigg(225-1140{M\over r}+
1400{M^2\over r^2}+745{Q^2\over r^2}-1755{MQ^2\over r^3}+
528{Q^4\over r^4}\bigg)
\nonumber\\
&&-{\beta^3 Q^2\over 710\pi r^{10}}\bigg(176400-1542240{M\over r}+
4294080{M^2\over r^2}+1123920{Q^2\over r^2}-5993680{MQ^2\over r^3}
\nonumber\\
&&-3841600{M^3\over r^3}+2005316{Q^4\over r^4}+7772240{M^2Q^2\over r^4}-
5069085{MQ^4\over r^5}+1074388{Q^6\over r^6}\bigg)\nonumber\\
&&+{\cal O}(\beta^4)~.\label{ttt}
\end{eqnarray}

Since the effective energy-momentum tensor keeps its property of
being traceless, the other two non-vanishing components can be
deduced from the two above, i.e.

\begin{equation}
T_\vartheta^{{\rm eff}~\vartheta}=T_\varphi^{{\rm eff}~\varphi}=
-{1\over2}(T_t^{{\rm eff}~t}+T_r^{{\rm eff}~r})~,\label{tff}
\end{equation}

Replacing this into Eqs.\ (\ref{3.21})--(\ref{gtt}) we obtain the metric
components
\begin{eqnarray}
g_{rr}^{-1}(r)&=&1-{2M\over r}+{Q^2\over r^2}-{2\beta Q^2\over 5r^4}
\bigg(10-15{M\over r}+6{Q^2\over r^2}\bigg)\nonumber\\
&&+{\beta^2 Q^2\over 105r^{6}}\bigg(7560-31920{M\over r}
+33600{M^2\over r^2}+17880{Q^2\over r^2}-36855{MQ^2\over r^3}
+9856{Q^4\over r^4}\bigg)\nonumber\\
&&+{\beta^3 Q^2\over 5005r^{8}}
\bigg(14414400-110270160{M\over r}+
272912640{M^2\over r^2}+71431360{Q^2\over r^2}\nonumber\\
&&-342838496{MQ^2\over r^3}-
219739520{M^3\over r^3}+104276432{Q^4\over r^4}+
404156480{M^2Q^2\over r^4}\nonumber\\
&&-241626385{MQ^4\over r^5}+47273072{Q^6\over r^6}\bigg)+
{\cal O}(\beta^4)~,\label{3.22z}
\end{eqnarray}

\begin{eqnarray}
g_{tt}(r)&=&-\bigg(1-{2M\over r}+{Q^2\over r^2}\bigg)+
{2\beta Q^2\over 5r^4}\bigg( 5 -5{M\over r}+{Q^2\over r^2}\bigg)
\nonumber\\
&&-{\beta^2 Q^2\over 105r^{6}}\bigg(2520-8400{M\over r}
+6720{M^2\over r^2}+3600{Q^2\over r^2}
-5355{MQ^2\over r^3}+952{Q^4\over r^4}\bigg)
\nonumber\\
&&+{\beta^3 Q^2\over r^{8}}
\bigg(-720+4656{M\over r}-9600{M^2\over r^2}-
{13504\over 5}{Q^2\over r^2}\nonumber\\
&&+{592864\over55}{MQ^2\over r^3}+6272{M^3\over r^3}+{3403504\over1155}
{Q^4\over r^4}+{112064\over11}{M^2Q^2\over r^4}\nonumber\\
&&+{2295703\over429}{MQ^4\over r^5}-{1235296\over1365}{Q^6\over r^6}
+{64\over91}{M^2Q^4\over r^6}+{\cal O}(1/r^7)\bigg)+
{\cal O}(\beta^4)~.\label{3.22}
\end{eqnarray}

Since the complete expression for $g_{tt}$ is already rather complicated
we do not pursue the computation of higher order corrections.

The validity of this metric will be assured if the condition
Eq.\ (\ref{2.2})
holds. In our case, this takes the form $-\beta Q^2/r^4\ll 1$. Thus,
Eqs.\ (\ref{3.22})--(\ref{3.22z})
will be a good approximation to the (even extremely) charged black hole
solutions in quadratic theories if
\begin{equation}
r_H\gg\sqrt{-\beta}~, \label{3.23}
\end{equation}
where $r_H$ is the radial coordinate of the event horizon (in
Schwarzschild's gauge).

The radial coordinate of the horizon can be computed from
making vanish $g_{tt}(r_H)$ given by Eq.\ (\ref{3.22}).
To first order in $\beta$ we have (for non-extreme black holes)
\begin{eqnarray}
r_H&=&r_+ +{\beta Q^2\left(5r^2_+-3Q^2\right)\over
5r^3_+ \left(r^2_+-Q^2\right)}-{\beta^2 Q^4\over 1050r^7_+ }
{\left(2925r^6_+-6515Q^2r^4_++5095Q^4r^2_+-1337Q^6\right)\over
\left(r^2_+-Q^2\right)^3}\nonumber\\
&-&{\beta^3 Q^2\over 750750r^{11}_+ \left(r^2_+-Q^2\right)^5}
\bigg(600600r^{14}_+-43507050Q^2r^{12}_++75284175Q^4r^{10}_+
-79088475Q^6r^8_+\nonumber\\
&+&36916630Q^8r^6_+-435320Q^{10}r^4_+-
5714693Q^{12}r^2_++1442637Q^{14}\bigg)
+{\cal O}(\beta^4)\label{3.25}~,
\end{eqnarray}
where $r_+=M+\sqrt{M^2-Q^2}$.

We observe that the quadratic black hole shrinks with respect to the
corresponding general relativistic one when we consider up to
$\beta^2$ terms, but the next order already tends to stabilize the
solution near the general relativistic value.
In fact, we can expect $r_H>0$ for a wide range of
$\beta$, thus ensuring the existence of charged black holes in quadratic
theories of gravitation.

The extreme black hole will be now reached with a maximal charge
different from the general relativistic one
\begin{equation}
Q^2_{\text{extr}}=
M^2+{2\over 5}\beta+{4\over 525}{\beta^2\over M^2}
+{\cal O}(\beta^3)~,\label{3.25b}
\end{equation}
thus $r_H$ given by Eq.\ (\ref{3.25}), will remain always bounded.
In fact,
\begin{equation}
r^{\text{extr}}_H=M+{\beta\over5M}
+{\cal O}(\beta^2)~.\label{3.25c}
\end{equation}
[Note that since $g_{tt}\sim (r-r^{\text{extr}}_H)^2$ for an extreme
black hole to obtain $r^{\text{extr}}_H$ to order $\beta^n$ we need
to know $g_{tt}$ to order $\beta^{2n}$].

The horizon area, $A_H$, is given by
\begin{eqnarray}
A_H&=&4\pi r^2_H=
4\pi r^2_+ + {8\pi\beta Q^2\over 5r_+^2\left(Q^2-r_+^2\right)}
\left(3Q^2-5r_+^2\right)\nonumber\\
&&+{16\pi\beta^2 Q^4\over 525r_+^6\left(Q^2-r_+^2\right)^3}
\left(-287Q^6+1069Q^4r_+^2-1340Q^2r_+^4+600r_+^6\right)\nonumber\\
&&+{8\pi\beta^3Q^2\over 375375r_+^{10}\left(Q^2-r_+^2\right)^5}
\bigg(434532Q^{14}-999705Q^{12}r_+^2-5007445Q^{10}r_+^{4}
+24633770Q^8r_+^6\nonumber\\
&&-43546450Q^6r_+^8+38687775Q^4r_+^{10}-17253525Q^2r_+^{12}
+3003000r_+^{14}\bigg)+{\cal O}(\beta^4)~.
\label{3.26}
\end{eqnarray}

\section{Thermodynamic properties of the perturbative solution}

In the theory of black holes in classical general relativity a
mathematical analogy was discovered between certain laws of black
hole mechanics and the ordinary laws of thermodynamics. With
the discovery of quantum particle creation near black holes,
this analogy appeared to be something more than a simple mathematical
analogy.

Very recently, some aspects of this significative relation
between black hole mechanics and thermodynamics have been studied
in a more general context than in Einstein's general relativity.
In particular, here, we want to know for our approximate solution
what is the explicit form of the laws of black hole mechanics.

As it is known the {\it zeroth} law of black hole mechanics asserts
that the surface gravity $\kappa$ (where $\kappa$ is defined by
$\xi^b\nabla_b\xi^a=\kappa\xi^a$ with $\xi^a$ is the null generator of
horizon) is constant all over the event horizon of a stationary black hole.
The proof of this law makes direct use of the specific form of
Einstein field equations and matter dominant conditions \cite{BCH73},
and it was not extended to others theories of gravity. Anyway, it
can be easily shown that the constancy of the surface gravity trivially
holds in the spherically symmetric case and can be computed using
Eq.\ (\ref{3.24}) below.

In fact, we can safely compute the Bekenstein-Hawking temperature from
the surface gravity, $\kappa$, since this equation was found
euclideanizing the black hole metric without making any assumption on
the field equations that it fulfilled \cite{HH76},
\begin{equation}
T_H={\kappa\over 2\pi}=-{1\over 4\pi}{g'_{tt}\over\sqrt{-g_{tt}g_{rr}}}
\biggr\vert _{r=r_H}~,\label{3.24}
\end{equation}
where $g'_{tt}=\partial{g_{tt}}/\partial r$.

Thus, in our approximation, for non-extreme black holes we have,
\begin{eqnarray}
T_H&=&{1\over4\pi r_+^3}\left(r^2_+-Q^2\right)+
\beta {Q^4\over 5\pi r^7_+\left(r^2_+-
Q^2\right)} \left(2r^2_+-Q^2\right)
\nonumber\\&&+\beta^2{Q^4\over 525\pi r_+^{11}\left(Q^2-r_+^2\right)^3}
\left(168Q^8-676Q^6r_+^2+791Q^4r_+^4-100Q^2r_+^6-225r_+^8\right)
\nonumber\\
&&+\beta^3{Q^2\over 3003000\pi r_+^{17}\left(r_+^2-Q^2\right)^5}
\bigg(-1848000Q^{18}-46018351687 Q^{16}r_+^2\nonumber\\
&&+85649730010 Q^{14}r_+^{4}+215820442547 Q^{12}r_+^6
-752950763928 Q^{10}r_+^8\nonumber\\
&&+751221815775 Q^8r_+^{10}-212346839950 Q^6r_+^{12}
-88152761275 Q^4r_+^{14}\nonumber\\
&&+46904894700 Q^2r_+^{16}-126126000 r_+^{18}+{\cal O}(1/r_+)\bigg)
+{\cal O}(\beta^4)~,\label{3.28}
\end{eqnarray}

Note that, since $\beta<0$, the effect of the quadratic
gravitational theory corrections at first order in $\beta$
will be that of decreasing the
black hole radiation temperature with respect to the general
relativistic value (with the same $M$ and $Q$),
leaving thus out open the possibility of switching off black
hole evaporation and leaving behind a charged remnant with a mass of the
order of the Planck mass.
Higher order corrections, however, alternate sign. To further discuss
this possibility one should have an exact solution and stablish the values
of $\beta$ that allow switching off of the black hole temperature. From
Eq.\ (\ref{3.25b}) and Eq.\ (\ref{3.25c}) we can check that
$T_H(Q_{max})=0$ up to order $\beta$.

The question of the {\it first} law of black hole mechanics in
a general theory of gravity was largely discussed in Refs. \cite{SW92,W93}
and clearly holds in our case, since the quadratic theory is derived from
the diffeomorphism invariant lagrangian (\ref{1.1}).
Inverting and integrating the fundamental relation
\begin{equation}
dM={\partial M\over \partial A}\biggr\vert _Q dA+{\partial M\over
\partial Q}\biggr\vert _A dQ=T_HdS+\Phi_HdQ~, \label{3.29}
\end{equation}
in Ref. \cite{CLA94}, we found, to first order in $\beta$, an
approximate expression for the entropy of the charged black hole, in
which clearly appears not simply one quarter of the
area of the horizon, but in general a more complicated function of the area,
$A$ and charge, $Q$.

Recently, in Ref. \cite{W93}, Wald has developed a
mathematical rigorous method to derive
the exact formal expression of the entropy of a stationary black hole
in any general covariant theory of gravity. In fact, the entropy was
found to be a local geometrical quantity expressed as an
integral evaluated on an arbitrary cross-section of a killing
horizon over the associated Noether charge.

In Ref. \cite{JKM94} Wald's Noether charge techniques have been explicitly
applied to a lagrangian theory such as (\ref{1.1}), which is
clearly generally covariant (invariant under general diffeomorphism
transformations).
Their general expression for the entropy can be written as
\begin{equation}
S={1\over 4}\int_{\Sigma}{d^2x\sqrt{h}\big[1+2\alpha
R+\beta g_{\bot}^{\mu\nu}R_{\mu\nu}\big]}~,\label{3.30}
\end{equation}
where $\Sigma$ is any arbitrary cross-section of the horizon,
$d^2x\sqrt{h}$ is the intrinsic volume element and where $g_{\bot}^{\mu\nu}=
g^{\mu\nu}-h^{\mu\nu}=(\xi^\mu\chi^\nu+\chi^\mu\xi^\nu)$
is the metric in the subspace normal to the horizon, with
$\xi^\mu$ the stationary killing field and $\chi^\mu$ a vector
field orthogonal $\Sigma$ satisfying $\chi^\rho\chi_\rho=0$
and $\chi^\rho\xi_\rho=1$ on the horizon.

In fact, making use of expression (\ref{3.30}), the entropy
for our approximate solution takes the following form
\begin{eqnarray}
S&=&{1\over 4}\int_{r=r_H}{d\vartheta d\varphi\sqrt{g_{\vartheta\vartheta}
g_{\varphi\varphi}}\big[1+2\beta g^{tt}R_{tt}\big]}
={1\over 4}\left[1+16\pi\beta T_t^{{\rm eff}~t}\right]_{r=r_H}\times
\int_{r=r_H}{d\vartheta d\varphi\sqrt{g_{\vartheta\vartheta}
g_{\varphi\varphi}}}\nonumber\\
&=&{A\over 4}-8\pi^2\beta{Q^2\over A}
-1024\pi^4\beta^3{Q^2\over A^5}(A-4\pi Q^2)^2
+16384\pi^5\beta^4{Q^2\over 5A^7}(A-4\pi Q^2)^{-1}\times\nonumber\\
&&(25A^4-470\pi Q^2A^3+3506\pi^2Q^4A^2-11120\pi^3Q^6A+12448\pi^4Q^8)
+{\cal O}(\beta^5)~,\label{3.31}
\end{eqnarray}
where we have taken the bifurcation surface as the cross-section of the
horizon $\Sigma$ and where, in the charged spherically symmetric case,
$R=0$ and $g_{\bot}^{tt}=2/g_{tt}$ is the only non vanishing
component of the metric in the subspace normal to the horizon.
In addition, Eq.\ (\ref{ttt}) was used for computing the
term $g^{tt}R_{tt}=8\pi T_t^{{\rm eff}~t}$, and the area of the event
horizon is given by Eq. (\ref{3.26}).

Note that the knowledge of the metric to ${\cal O}(\beta^3)$ allowed us
to compute the entropy up to ${\cal O}(\beta^4)$. That means that from
the expression of the Reissner--Nordstr\"om metric we can compute $S$ to
${\cal O}(\beta)$, and thus recover the results of Refs. \cite{AEL93,EL94}
without having to solve any further field equation for $g_{\mu\nu}$.
The vanishing of the term proportional to $\beta^2$ is a casual fact due
to the particular dependence of $T_t^{{\rm eff}~t}$ to ${\cal O}(\beta)$
(that order vanishes when evaluated on the horizon).

In Ref. \cite{JKM94} it was noted that, since the action
Eq.\ (\ref{1.1}) ``is linear in $\alpha$ and in $\beta$, the
modifications to the entropy from higher curvature terms in the
original action would only be linear in $\alpha$ and in $\beta$".
The precise meaning of this statement
could not appear clear to the reader at first sight.
When we write explicitly the entropy as a function of the horizon
area and charge, we find corrections coming from quadratic
curvature terms in the general formula Eq.\ (\ref{3.30}) that are
not linear, but of $(n+1)$-th order in $\beta$, if the metric contains
$n$-th order terms in this coupling constant.

To continue our previous discussion on the validity of the laws of
black hole mechanics in the case of our approximate solution, we consider
the the {\it second} law of thermodynamics,
which states that the entropy must be an always increasing function.
The theorem that establishes that the
area of the black hole horizon must be an always increasing quantity,
was proved only for general relativity \cite{H73}.

Although any proof of the second law of black
hole mechanics in a general theory of gravity has not been given yet,
it seems to exist a relation between its validity and
the positive energy condition of the theory.
In the case of lagrangian theory examined in this paper (Eq.\
(\ref{1.1})), the positive energy condition does not holds in general,
as we can see from the linearized field equations given in Ref.
\cite{AEL93}, where the presence of the spin two massive tensorial
field $\psi^{ab}$ allows for negative values of the energy conditions on
the energy-momentum tensor.

As in general relativity, here, the {\it third} law of thermodynamics
should hold in the form due Nerst: Loosely speaking, ``it is very hard to
reach $T_H=0$". The version due to Planck does not hold, since from Eqs.
\ (\ref{3.25b})-(\ref{3.25c}) we have
\begin{equation}
S(T_H=0)=\pi M^2-{8\pi\over 5}\beta+{\cal O}(\beta^2/M^2).
\end{equation}
{}~~This ground state is clearly degenerate due to the general relativistic
term. The contribution of
quadratic theories is (to first order) only
to contribute with an universal constant (function of $\beta$).

With regards to the {\it fourth} law or the scaling laws fulfilled by the
black hole critical exponents\cite{L93,L94}, we expect them to hold also
for our approximate solution. In particular, the black hole critical
exponents should be the same as in general relativity, since the
Reissner--Nordstr\"om solution gives the leading behavior and since
the validity of the universality hypothesis\cite{L94}.

\section{Discussion}

The aim of this work was two-fold: First to see if, in practice, our
perturbative approach allowed to be carried out to orders higher than
the first.
This has been shown to be possible explicitly up to third order.
The second objective of this work was to see if from the higher orders
one could guess some recursive formulae that would lead us to an {\it
exact} solution.
So far, we have not fulfilled this expectation, but may be the
reader can do it. We can, however, try to say something about the
$\beta$--gravity (or $\beta$--sector of the quadratic theories):
Although the first order corrections
seem to indicate that $\beta$--gravity is
weaker than General Relativity, we can see
(for example in $Q_{\text{max}}$, $r_H$ and $T_{H}$) that to higher
order we have alternateness in the sign of the
corrections (we needed to go up to 3rd. order in $\beta$ to check that).
Using the Noether charge method we have been able to express the black hole
entropy in terms of the horizon area and charge up to 4th. order in $\beta$,
and (in spite of the casual vanishing of the ${\cal O}(\beta^2)$ term) we
have shown that the dependence of $S$ on $\beta$ is not simply linear.
We have also studied the extremely charged black hole. There, due
to the particularity of this solution, to obtain
$r_H^{\text{extr}}$ (and the other quantities evaluated on the extreme
horizon such as $T_H$ and $S$) to the first order in $\beta$ we had to
compute $g_{\mu\nu}$ to order $\beta^2$.\par\noindent
{}~~From the fact that our perturbative method can not go
to the high curvature regime, unfortunately, we cannot say anything
about the most interesting issue of the singularity at $r=0$.\par\noindent
{}~~We are presently dealing with the rotation effects, i.e.
the $\beta$ first--order corrections to the Kerr--Newman
solution and the results will be published elsewhere\cite{L95}.

\begin{acknowledgments}
This work was partially supported by the Directorate General for
Science, Research and Development of the Commission of the European
Community. C.O.L was supported by the Direcci\'on General de
Investigaci\'on Cient\'\i fica y T\'ecnica of the Ministerio de
Educaci\'on
y Ciencia de Espa\~na and CICYT AEN 93-0474. M.C. holds an
scholarship from the Deutscher Akademischer Austauschdienst.
\end{acknowledgments}

\end{document}